\documentclass[12pt,preprint]{aastex}
\usepackage[usenames]{color}

\shorttitle{Laboratory Experiment of Magnetic Reconnection in Chromospheric Jets}
\shortauthors{Nishizuka et al.}

\begin{document}

\title{A Laboratory Experiment of Magnetic Reconnection: Outflows, Heating and Waves in Chromospheric Jets}


\author{N. Nishizuka\altaffilmark{1}, Y. Hayashi\altaffilmark{2}, H. Tanabe\altaffilmark{2}, A. Kuwahata\altaffilmark{2}, Y. Kaminou\altaffilmark{2}, 
Y. Ono\altaffilmark{2}, M. Inomoto\altaffilmark{2}, T. Shimizu\altaffilmark{1}
}

\altaffiltext{1}{Institute of Space and Astronautical Science, Japan Aerospace Exploration Agency, Yoshinodai, 
Sagamihara, Kanagawa 229-8510, Japan; nishizuka.naoto@jaxa.jp}
\altaffiltext{2}{Graduate School of Frontier Sciences, University of Tokyo, Tokyo, Japan}


\begin{abstract}
Hinode observations have revealed intermittent recurrent plasma ejections/jets in the chromosphere. These are interpreted as a result of non-perfectly anti-parallel magnetic reconnection, i.e. 
component reconnection, between a twisted magnetic flux tube and the pre-existing coronal/chromospheric magnetic field, though the fundamental physics of component reconnection is unrevealed. 
In this paper, we experimentally reproduced the magnetic configuration and investigated the dynamics of plasma ejections, heating and wave generation triggered by component reconnection in the 
chromosphere. We set plasma parameters as in the chromosphere (density 10$^{14}$ cm$^{-3}$, temperature 5-10 eV, i.e. (5-10)$\times$10$^4$ K, and reconnection magnetic field 200 G) using 
argon plasma. Our experiment shows bi-directional outflows with the speed of 5 km s$^{-1}$ at maximum, ion heating in the downstream area over 30 eV and magnetic fluctuations mainly at 5-10 
$\mu$s period. We succeeded in qualitatively reproducing chromospheric jets, but quantitatively we still have some differences between observations and experiments such as jet velocity, total 
energy and wave frequency. Some of them can be explained by the scale gap between solar and laboratory plasma, while the others probably by the difference of microscopy and macroscopy, 
collisionality and the degree of ionization, which have not been achieved in our experiment.
\end{abstract}


\clearpage
%
\section{Introduction}
Solar jets are ubiquitous phenomena both in the corona and the chromosphere. They sometimes show cusp-like structure which is thought to be evidence of magnetic reconnection and it is 
interpreted that reconnection between an emerging flux rope and the pre-existing coronal/chromospheric field may ubiquitously produce solar jets \citep[e.g.][]{shi94, yok95}. Recently, the Solar 
Optical Telescope \citep[SOT;][]{tsu08,sue08, ich08, shim08} on board Hinode \citep{kos07} acquired a time series of Ca II H images in high spatial resolution,  which revealed the fine structure of 
the sunspot magnetic field and the dynamic activities in the chromosphere, such as penumbral jets \citep{kat07a}, chromospheric anemone jets \citep{shi07, nis08, nis11, sin11} and recurrent plasma 
ejections over the sunspot light bridge \citep{shim09, shim11}. The vector magnetogram observations with Hinode also revealed that even component magnetic reconnection, i.e. reconnection between 
non perfectly anti-parallel magnetic field lines, can be origins of jet activities for the first time. Especially in solar phenomena observed as evidence of component reconnection, chromospheric plasma 
ejections from sunspot light bridges are relatively well revealed with their photospheric magnetic configurations. These show current-carrying highly twisted magnetic flux tubes trapped along the light 
bridge below a canopy structure of the umbral fields, forming the magnetic configuration preferable for magnetic reconnection at low altitudes \citep{shim09}. However it is unrevealed what differs 
magnetic reconnection occurring in this magnetic configuration from the perfectly anti-parallel magnetic reconnection and whether it is possible to generate dynamic ejections as observed by this 
magnetic configuration. 

Figure 1(a) shows the general magnetic configuration of solar jets originating magnetic reconnection between an emerging flux tube and the pre-existing vertical coronal or chromospheric magnetic 
field, as well as in sunspot light bridges. An emerging flux tube has a helical magnetic structure and interacts with the surrounding vertical field lines when it is lifted up by magnetic buoyancy. Two 
non-perfectly anti-parallel magnetic field lines of the emerging flux rope and the ambient corona/chromosphere reconnect at the one side of the flux rope. This process has been modeled by 3-dimensional 
magnetohydrodynamic simulations \citep{mor08, par10, mag10}. The helical magnetic configuration in the laboratory is called spheromak, which has toroidal and poloidal components of magnetic field 
forming twisted magnetic flux tube as well as the emerging flux rope in the solar atmosphere. Here we note that toroidal field is directed along the flux tube and poloidal field circulates around the 
toroidal field. Figure 1(b) shows the illustration of magnetic reconnection between the spheromak and the newly emerged straight field lines induced by the center solenoid coil. Also in this case, 
helical field lines of the spheromak and the newly emerged straight field lines cross to cause component reconnection at the contact surface of the two.  

Hinode observations have revealed high spatial magnetic configurations of chromospheric plasma ejections but they still have limits to reveal the fundamental physics of magnetic reconnection in 
much smaller spatial scale and its 3-dimensional structure, because the spatial resolution of SOT/Hinode is 150 km on the solar surface and only a single layer of the chromosphere is observable. 
Hinode also has difficulties to observe real plasma flows with Doppler measurements, ion/electron temperatures, electrical resistivity and magnetic fluctuations, as well as magnetohydrodynamic 
simulations of jets. On the other hand, laboratory plasma experiment has an advantage to directly measure the plasma condition. In this paper, we reproduced magnetic configuration of solar jets 
driven by an emerging flux rope and component reconnection, by focusing on the similarity of the emerging flux and the spheromak. Then we investigated the details of the plasma ejections with 
direct measurements in the laboratory plasma experiment.

In \S2, we explain the apparatus of the reconnection experiment of laboratory plasma and the method of measurements. In \S3, we show experimental results such as reconnection jets, plasma 
heating and wave generation. Finally, in \S4, we discuss the experimental results comparing with solar observations of chromospheric jets.

%
\section{The TS-4 Experiment}
\subsection{Apparatus of the TS-4 Experiment}

The Tokyo Spheromak (TS) toroidal plasma merging experiments have been conducted to study plasma heating effects of magnetic reconnection since 1986 \citep[e.g.][]{ono93, ono97, yam97}. As 
shown in Figure 2, the instrument of TS-4 device is composed of the axisymmetric toroidal vacuum vessel with length of 2.5 m and diameter of 2.0 m and the two flux cores of poloidal field (PF) and 
toroidal field (TF) coils for poloidal/toroidal flux injections \citep[see more details in][]{kaw05, kaw07}. The cylindrical coordinate is adopted: Z-axis in the axial direction, R-axis in the radial direction 
and T-axis in the toroidal direction. TS-4 also contains a center solenoid coil called OH coil (originally named ohmic heating coil) along the Z-axis with the radius of 90 mm, outside of which is filled 
with fully ionized plasma and inserted by three kinds of probe arrays: the 2D magnetic field probe array, the magnetic fluctuation probe array and the Mach probe. Locations of the probe arrays are 
overlaid in Figure 2. The vacuum vessel has a window, through which the Doppler spectroscopy measurements are performed. With these instruments, we performed four different measurements 
simultaneously for the plasma diagnostics in TS-4 device. 

Figure 3 shows the schematic pictures of our experimental scenario. The time profiles of electrical current induced in TF, PF and OH coils are shown in Figure 4. Initially the vacuum vessel is kept in 
the vacuum state less than 10$^{-6}$ Torr and then filled with argon gas. The reasons for the selection of argon gas are reproductivity and emissivity compared with hydrogen. At first, argon plasma 
was fully ionized (more than 90\%) and two toroidal plasmas with the radius of 0.5 m were generated by the flux cores (Fig. 3a). Two toroidal plasmas were merged together in the axial (Z) direction 
under magnetic compression provided by the pair of PF coils or acceleration coils (Fig. 3b). The magnetic reconnection occurs at the contacting point of the two toroidal plasmas, causing high-power 
plasma heating by magnetic reconnection. The merged toroidal plasma, i.e. spheromak, has the helical magnetic configuration with toroidal and poloidal components. Then we induced a new poloidal 
field anti-parallel to the poloidal field of the spheromak by the OH coil (Fig. 3c). The spheromak and the anti-parallel OH field were merged together in the radial (R) direction via magnetic reconnection. 
As the reconnection proceeds, the spheromak becomes smaller and smaller and the reconnection point moves upward (in positive $R$-direction) under control by the separation coils (Fig. 3d).

The experimental parameters were selected similar to the chromospheric parameters in the sun such as ion and electron temperatures $T_i$=$T_e$=10 eV ($\sim$10$^5$ K), electron density $n_e 
\sim$10$^{14}$ cm$^{-3}$, and toroidal and poloidal magnetic fields are B$_p$/B$_t$= 400G/220G at the surface of the spheromak (plasma beta $\beta\sim$0.06-0.24). Here we note that, with the 
above experimental parameters, ion cyclotron frequency and gyro period are 50-150 kHz and 6-20 $\mu$s (100-300 G), ion Larmor radius 7 cm, local sound speed 7 km s$^{-1}$ (T=10$^5$ K) and 
ion Alfv\'{e}n velocity 13 km s$^{-1}$ (B=200 G) for argon ion we used, respectively. Therefore, in this experiment, we see microscopic phenomena around the reconnection region occurring in the 
solar chromosphere, which cannot be seen neither in Hinode observations nor in magnetohydrodynamic simulations. 

\subsection{TS-4 Diagnostics}

The two sets of the 10$\times$9 arrays of magnetic pickup coils were inserted in the $R-Z$ plane of the vessel to measure directly the 2D magnetic field profile. Its spatial resolution is 80 mm in the 
radial direction and 90 or 135 mm in the axial direction. Its time resolution is 1 $\mu$s. The poloidal flux contours and current density profiles, based on the axial symmetry assumption, were calculated 
from the measured $B_z$ and $B_t$ components of 2D magnetic field profiles. The current sheet was identified by the measured toroidal current density profile $J_t$ and the X-line structure. More in 
detail, physical parameters of $B_r$, $J_t$ and $E_t$ were calculated from $B_z$ by the following equations,
\begin{eqnarray}
B_r(r,z) & = & -\frac{1}{2\pi r}\frac{\partial \Psi}{\partial z}\\
J_t(r,z) & = & \frac{\partial B_r}{\partial z}-\frac{\partial B_z}{\partial r} = -\frac{1}{2\pi r}\frac{\partial^2 \Psi}{\partial z^2}-\frac{\partial B_z}{\partial r} \\
E_t(r,z) & = & -\frac{1}{2\pi r}\frac{\partial \Psi}{\partial t},
\end{eqnarray}
where $\Psi(r,z,t) = \int^r_{r_{min}} 2\pi r' B_z dr'$ is magnetic flux ($r_{min}$=92 mm at the surface of the OH coil). Since magnetic field is almost negligible in the current sheet, the effective resistivity 
is derived by $\eta$=E$_t$/J$_t$, although it is slightly affected by the magnetic flux inside the OH coil.

An array of 27 magnetic pickup coils \citep{kuw11} was also inserted on the mid-plane (z=0), but with much shorter distances of 10 mm, to measure the 1D magnetic fluctuation profile in R-direction 
($\delta B_z$). The magnetic fluctuation probe array is digitized by 200 M sampling s$^{-1}$ with the 8 bit analog to digital converters (ADCs). The integration of measured $\delta$B$_z$ at the same 
location gives us $B_z$ value, and the toroidal current density is calculated by J$_t$= -$\partial$B$_z$/$\partial$r, on the assumption that $B_r$ is negligible on the mid-plane.

The ion Doppler velocity $V_i$ and temperature $T_i$ diagnostics was performed by the fiber optic multi-channel imaging spectroscopy system \citep{tan10}. 23 sets of multi-chord line spectra of Ar II 
at 434.8 nm are collected to optical fibers. Each of them are collimated by f=50 mm and F=1.2 camera lens and measured by ICCD imaging spectrometer with 256 pixel wavelength resolution (0.024 nm 
pixel$^{-1}$). At first, each Doppler width of the spectral line is calculated by using Gaussian function fitting algorithm to plot the 1D profile of $V_i$ and $T_i$ along the axial (Z) direction in the current 
sheet. Its spatial resolution is 20 mm in the axial direction, and its time resolution is 20 $\mu$s. 

The 1D ion velocity profile $V_i$ was also directly measured by the Mach probe array inserted along the current sheet near the OH coil at $r$=190 mm. Its axial spatial resolution is 50 mm and its time 
resolution is 0.5 $\mu$s. The probe was used to directly measure Mach number of the ion flow, which is calculated from the difference of the two current densities entering into the probe through the 
sheath, such as exp($KM$) = $J_{up}$/$J_{down}$. Since a paper on the K-value calibration for the Mach probe concluded K=2.0-2.5 for SSX parameters \citep[T$_e$=7 eV, T$_i$=20 eV][]{zha11}, we 
adopted K=2.5 in this paper.

%
\section{Experimental Results}

Figures 5(a) and 5(b) show the poloidal flux contours in the poloidal (R-Z) plane of two merging toroidal plasmas (B$_p$=400 G, B$_t$=220 G at the surface of the spheromak). These two toroidal plasmas 
were merged together during 380-400 $\mu$s (Fig. 5a and illustrations in Figs. 3a-3b). Magnetic reconnection between the spheromak and the OH field occurred during 450-560 $\mu$s (Fig. 5b and 
Figs.3c-3d). At that time, negative J$_t$ region (blue color in Fig. 5b), that is a current sheet, was observed between the spheromak and the OH coil (92 mm$<$r$<$160 mm), and simultaneously was 
the effective resistivity enhancement observed (discussion in $\S$4.2 and Fig. 12). As shown in Figure 5(c) of the radial (R) profile of B$_z$ component measured by the magnetic fluctuation probe array, 
the reconnection point between the spheromak and the OH field moved upward as the spheromak became smaller and smaller. The reconnection point is located in the undetectable area below the wall 
of the vacuum vessel in the early phase, where reconnection occur partially in vacuum and partially in plasma leading to the morphology change of the global magnetic field to accelerate the plasma by 
sling-shot effect due to magnetic tension force. In the later phase, we directly observed the reconnection point with the 2D magnetic probe array, when it moved upward as reconnection proceeds. 

Figure 6 shows the axial (Z) profiles of ion velocities during 380-560 $\mu$s measured by the fiber optic multi-channel imaging spectroscopy system. To distinguish 2-dimensional flow from the 
integrated line spectra in the line-of-sight direction, the system has 3 different views, from left-hand side, right-hand side and just above the mid-plane. Their line-of-sight directions are shown in 
Figure 6(a). Since they are set to avoid the center solenoid coil, each spectrum contain toroidal velocity, but it is less than 3 km s$^{-1}$ estimated from the bias velocity in view 3 and removed. Finally, 
the residual data  of view 1 and 2 in Figure 6(b) show outflow velocities in the poloidal (R-Z) plane stereoscopically. Positive Doppler values mean ion flow in the direction opposite to each camera lens. 
Initially plasma velocity was 0 km s$^{-1}$. Reconnection outflow was detected for 80 $\mu$s after 480 $\mu$s, during which ion flow was gradually accelerated to 3-4 km s$^{-1}$ in the line-of-sight 
direction (almost positive R-direction) at the fiber channel numbers 2, 7 and 13 in Figure 6(a) (near z=$\pm$400 mm). The bi-directional outflow was observed at around the fiber channel numbers 4 
and 10, which may correspond to the reconnection X-point. The reconnection outflow was accelerated in time and with distance from the X-point. 

Figure 7 shows the axial (Z) profiles of ion velocity V$_i$ in Z-direction during 380-560 $\mu$s, directly measured by the Mach probe at six positions $z$=100, 150, 200, 250, 300, 400 mm by turns. 
This is a complementary measurement for the previous Doppler spectroscopy. The ion velocity is presented in Mach number (local ion sound speed C$_s \sim$7 km s$^{-1}$ for $T_i$=10 eV). Before 
the reconnection, the ion velocity is detected to be zero. Positive velocity means rightward flow (outflow) from the mid-plane or the X-point. A reconnection X-point, i.e. the zero velocity point, 
existed between z=100-170 mm before 460 $\mu$s and then moved inward less than z=100 mm as reconnection proceeds. The maximum velocity was 0.4 $C_s$ (sound speed) at z=200 mm. Inside 
the velocity peak, ion flow was accelerated proportional to the distance from the X-point. Outside z=200 mm, ion velocity gradually decreased. These are consistent with the previous Doppler 
spectroscopy measurement. 

Additionally we derived ion temperature from the Doppler spectroscopy measurements. Figure 6(c) shows the axial (Z) profile of the 1D ion temperature T$_i$ during the spheromak and the OH field 
merging (420-560 $\mu$s). The initial ion temperature was uniformly 5-10 eV. The ion temperature at the left downstream area was preliminary heated up to 13-25 eV at first (400-440 $\mu$s). 
After that, the ion plasma was further heated up to 18-32 eV (440-480 $\mu$s) at maximum. It is gradually cooled down to 15-20 eV during 480-520 $\mu$s and then to 5-15 eV during 520-560 
$\mu$s. The locations of the highest ion temperature and the largest ion velocity are almost the same in the left downstream area in view 1, though the other side in view 1 and both in view 2 were 
not. 

Associated with magnetic reconnection, magnetic fluctuation of $B_z$ component ($\delta B_z$) was also observed by the magnetic fluctuation probe array at around the current sheet during 450-550 
$\mu$s. The locations of the fluctuation probe array and the corresponding 2D poloidal flux contours measured by the 2D magnetic probe array are shown in Figure 8(a). Figure 8(b) shows the time 
slice image of the magnetic fluctuations on the mid-plane, and Figure 8(c) shows three examples of $\delta$B$_z$ fluctuations. Since the current sheet and the reconnecting magnetic field lines 
are located in the axial (Z) direction, $\delta B_z$ fluctuations indicate longitudinal oscillation (magnetoacoustic sausage mode) or projected transverse oscillation with toroidal (guide) field to the 
poloidal (R-Z) plane. Here we can not distinguish standing waves from propagating waves with the current data. We applied wavelet analysis to these data. Details of the procedure are given by Torrence 
\& Compo (1998). Figure 9 displays the wavelet power spectra of magnetic fluctuations at different locations $r$=92, 112 and 182 mm on the mid-plane. In the wavelet spectrum diagrams, regions with 
95\% significance level are outlined. The power spectra show a peak in the period of 4-20 $\mu$s, in which there exists sub-structure at 5, 6, 10, 15-20 and 30 $\mu$s. The spectrum at 5 $\mu$s 
period also show two peaks at 440 $\mu$s and 480 $\mu$s in time variation (Fig. 9b). The oscillation lasts for 90 $\mu$s from 430 $\mu$s, containing 16 periods for the shortest frequency. Here we 
note that these oscillations are not affected by the magnetic fluctuations produced by the spheromak formation from 330-390 $\mu$s, because the detection times are completely different and the 
frequencies are slightly different from each other.

%
%
\section{Discussion and Conclusions}

We reproduced magnetic configuration of a twisted flux tube and chromospheric plasma ejections by component reconnection with laboratory experiment. Here we focused on the similarity between 
magnetic configurations of the spheromak in the laboratory plasma and of the emerging flux rope in the solar atmosphere. We performed two toroidal fully ionized argon plasmas merging experiment, 
followed by the magnetic reconnection driven by the OH field emergence. We measured 2D magnetic field configuration, ion flow, ion temperature and magnetic fluctuations at the same time during 
the reconnection process.

\subsection{Acceleration Mechanism of Jets}

Reconnection outflows were independently measured by the fiber optic multi-channel imaging spectroscopy system and the Mach probe. Both measurements show consistent results; the ion flow was 
accelerated proportional to the distance from the reconnection point and then decelerated by the accumulated magnetic flux at the outflow region. The maximum velocity in the line of sight direction 
(R-direction) $v_r$ was 4 km s$^{-1}$ by the Doppler measurement and the velocity in axial (Z) direction $v_z$ was 0.4  $C_s$ (sound speed) at maximum by the Mach probe, meaning 2.8 km s$^{-1}$ on 
the assumption that the local ion sound speed is 7 km s$^{-1}$. These values are about 40\% of the local Alfv\'{e}n velocity. In the later phase, plasma velocity near the mid-plane (z=100 mm) increased 
as well as in the outer region (Fig. 7b). At the same time, reconnection point, that is the transition of positive and negative (rightward and leftward) velocities, moved inward; it is located at z=150 mm 
at 420 $\mu$s and moved to z=100 mm at 460 $\mu$s and less than 100 mm later. This may suggest two possible interpretations: the one is that reconnection with a long current sheet transits to 
the X-type fast reconnection as shown in the illustration of Figure 10(a). The other one is that the reconnection point moves upward to the detectable region by the Mach probe, which is shown in 
Figure 10(b) for comparison. 

Here we note that the current sheet thickness is 10 cm, while ion skin depth and ion Larmor radius are 14-70 cm and 7-14 cm (20 eV, 20 mT), respectively. Therefore, the current sheet thickness is 
comparable to or a little smaller than the ion skin depth or the ion Larmor radius. However, the electron and ion mean free paths are 1 cm, which are smaller than the current sheet thickness, so that 
the Hall current is not detected in our experiment. The reconnection rate is estimated as $v_{in}$/$v_{out}$ =0.13-0.33 for $v_{in}$ =0.5-1.0 km s$^{-1}$ and $v_{out}$ =3-4 km s$^{-1}$, respectively, 
while the Sweet-Parker reconnection rate is 1/${R_m}^{1/2}$ =0.05-0.1 for $R_m$ =100-400. Hence, the reconnection in our experiment is slightly faster than the Sweet-Parker reconnection. Here 
we can say that collisionality suppresses the Hall effect and leads to the Sweet-Parker reconnection or slightly faster reconnection, though physical mechanism driving reconnection faster without 
the Hall effect is not revealed.

Generally, the plasma flow is accelerated by plasma pressure and magnetic tension force. However, no enhancements of temperature and density of ions and electrons were detected at the center of 
the current sheet in this experiment, so that it is expected that the plasma flow was not accelerated by plasma pressure but magnetically accelerated. Figure 11(a) shows the Lorentz force in Z-direction 
(J$\times$B)$_z$=J$_t$B$_r$. Near the mid-plane (-300 mm$<z<$300 mm), the Lorentz force plays a role in accelerating the plasma outward in the opposite directions from the mid-plane (z=0). Beyond 
300 mm apart from the mid-plane, the Lorentz force changes to the deceleration force due to the surrounding closed magnetic field. The absolute value of the Lorentz force increases in time and 
spatially in the axial and radial directions. If we assume the average value of the Lorentz force (J$\times$B)$_z$=10 N calculated from Figure 11(a), the acceleration is estimated 1.6$\times$10$^5$ 
km s$^{-2}$ and the accelerated velocity 15.9 km s$^{-1}$ with the acceleration time of 100 $\mu$s. This is comparable to but slightly larger than the measured values.

Next we show the moving velocity of the magnetic field lines at r=146 mm in Figures 11(b)-11(d), to compare with the measured ion flow. It is estimated from the Ohm's law with the 2D magnetic 
probe data ${\bf B}$=(B$_r$, B$_t$, B$_z$) and ${\bf E}$=(0, E$_t$, 0). The three components of the magnetic field velocity are given by
\begin{equation}
{\bf v}_{\perp} = \frac{{\bf E}\times {\bf B}}{|B|^2} = (\frac{E_tB_z}{|B|^2}, 0, -\frac{E_tB_r}{|B|^2}),
\end{equation}  
which is derived from the Ohm's law, ${\bf E}=-{\bf v}\times{\bf B}$. The z-component of $v_{\perp}$ in Figure 11(b) indicates the outward velocity of reconnected filed lines in the axial (Z) direction, 
where positive (negative) value means rightward (leftward) velocity. Similarly, the r-component of $v_{\perp}$ in Figure 11(c) indicates the inward velocity of the field lines to the current sheet in the 
radial (R) direction, where negative value means the inflow to the current sheet. Figure 11(d) shows the absolute value of the field line velocity. It shows the velocity peak at around 3-4 km s$^{-1}$ 
near z=200 mm. We can also find that the velocity peak moves inward from 420 $\mu$s to 540 $\mu$s, consistent with the Mach probe measurement of the ion flow, though argon gas is not 
completely frozen in to the magnetic field lines.

Our experimental results show the ion velocity with both axial and radial components, which means reconnection outflow is in the direction apart from the parallel direction to the surrounding stratified 
(vertical) magnetic field lines. This is consistent with the fact that magnetic tension force works obliquely to the stratified field with the angle of 45$^{\circ}$ as shown in Figure 5(b). However, in solar 
observations, it looks that plasma ejections over the light bridge occur along the vertical magnetic field lines. This is probably because the reconnection outflow is redirected to the parallel direction 
of the surrounding straight field lines after the collision of the plasma flow and the magnetic field lines. Therefore, even in the laboratory experiment, it would be reproduced in much larger spatial scale 
where magnetohydrodynamic behavior becomes dominant.

\subsection{Relationship among Jet, Heating and Wave Generation}

Figure 12(a) shows the relationship between the resistivity enhancement and the outflow acceleration in time. The effective resistivity is measured at r=119 mm on the mid-plane (z=0 mm) and peaks 
at 450-480 $\mu$s. The outflow velocities at two different locations z=200 mm and 300 mm are overlaid on it. The acceleration of reconnection outflow occurs just after the enhancement of effective 
resistivity. The peak of the outflow velocity at z=300 mm is earlier and smaller than the one at z=200 mm. Time variations of the ion temperature measured by the Doppler spectroscopy measurement 
with fiber channel 1, roughly at $z$=-300 mm and $r$=90-200 mm, and the power spectrum of magnetic fluctuations of 5 $\mu$s period at $r$=112 mm on the mid-plane (dotted line in Fig. 9b) are 
shown in Figure 12(b). The magnetic fluctuation shows three peaks at 445, 480 and 500 $\mu$s. The first one corresponds to the beginning of the resistivity enhancement, and the second one is at 
the peak time of the effective resistivity. 

Ion heating occurred in the left downstream area close to the surrounding closed magnetic field. The ion temperature increased to 20-30 eV. The heating by reconnection is expected T$_0$/$\beta$ 
in magnetohydrodynamic theory, where T$_0$ is the initial temperature and $\beta$ is the plasma beta \citep{yok95}. Assuming $\beta$ is 0.25, heating occurs from 10 eV to 40 eV. This is comparable 
to the measurement. The fact that ion heating occurs at the edge of the downstream area is consistent with the previous merging experiment \citep{ono11}, though the distance of the heating spot 
from the X-point is larger (approximately four times). This is probably because the distance from the X-point to the obstacle may determine the heating spot. Ono et al. (2011) suggested ion heating 
mechanisms by the fast shock and the viscosity in the downstream area, but the fast shock may not in this experiment because the measured Mach number is always less than unity. Rather it is 
interesting to see that the enhancement of magnetic resistivity and waves are in association with ion temperature enhancement. The locations of enhanced resistivity and hot ion spot is different 
from each other, so the association of ion heating and waves may indicate some physical relationship among them. In solar observations, it is impossible to detect $\mu$s order waves. However, this 
experimental result may suggest that such high frequency waves are generated through the magnetic reconnection process and contribute to the ion (and electron) heating.

Figures 13 shows the different magnetic configurations of the simple spheromak merging experiment and the OH field merging experiment. The former experiment confines hot plasma at the center 
of the spheromak with closed field lines, but the latter does not. In the OH field merging experiment, the hot plasma spot is not maintained for a long time and diffused along the surrounding straight 
open magnetic field lines. This would also affect the wave generation associated with magnetic reconnection. 

\subsection{Energy Estimation of Magnetic Reconnection}

Experimental measurements of ion temperature and velocity enable us to estimate thermal and kinetic energies converted from the released magnetic energy. Assuming plasma density $n=10^{14}$ 
cm$^{-3}$, the thermal energy is E$_{th}$=nk$_B$TV=1.1$\times$10$^7$ erg ($V$=3.8$\times$10$^3$ cm$^3$ and $T_i$=30 eV) and the kinetic energy is E$_{kin}$=Vnm$_i$v$_i^2$/2=3.2$\times$ 
10$^5$ erg (V=2.3$\times$10$^4$ cm$^3$ and $v_i$=4 km s$^{-1}$), leading to the total converted energy 1.1$\times$10$^7$ erg. It seems that kinetic energy is much smaller than thermal energy 
converted from magnetic energy in this experiment.

On the other hand, the reconnection rate of the total magnetic flux $\partial \Psi$/$\partial t$ is of the order of (2-3)$\times$10$^{10}$ Mx s$^{-1}$. By assuming two dimensional steady magnetic 
reconnection, the released magnetic energy can be estimated at the Poynting flux entering from both sides into the reconnecting region using the relation, 
\begin{equation}
\frac{dE_{mag}}{dt} =2\frac{B^2}{4\pi} v_{in} A,
\end{equation}
where dE$_{mag}$/dt is the magnetic energy release rate due to magnetic reconnection, $B$ is magnetic flux density in the spheromak, $v_{in}$ is an inflow velocity to the reconnection site, $A$ is 
the surface area of the current sheet ($A=2\pi rL_z$=2500 cm$^{2}$; for $r$=10 cm, $L_z$=40 cm). Since we cannot know the actual inflow velocity $v_{in}$, we use a perpendicular velocity to the 
magnetic field lines assuming zero resistivity outside the current sheet, such that ($v_{\perp}$)$_r$=$E_tB_r$/$|B^2|$ as an assumption. Figure 11(c) shows $(v_{\perp})_r$ profile in Z-direction at r=146 
mm, thus $(v_{\perp})_z$=0.5-1.0 km s$^{-1}$ and the resulting energy release rate d$E_{mag}$/dt=6$\times$10$^{12}$ erg s$^{-1}$, with the total energy release 3$\times$10$^8$ erg for the duration 
of magnetic reconnection, 50$\mu$s. These are comparable to the estimated total energy of thermal and kinetic energies 10$^7$ erg. 

As for the energy gap between laboratory experiments (10$^{8-9}$ erg) and solar observations (10$^{24}$ erg), it would be explained by the scale gap between laboratory and solar plasmas, because 
stored magnetic energy $E_{mag}\propto B^2L^3$ under the condition that B is constant. Since spatial scale of solar plasma is 4-5th orders larger than laboratory plasma, stored energy is self-similarly 
enlarged to $E_{mag,solar}$=10$^{12-15} E_{mag,lab} \sim$10$^{20-24}$ erg, corresponding to nanoflare energy regime of solar observations, that is, chromospheric jets. Similarly, reconnection time 
scale is determined by $\tau_{rec}$=$\sqrt{\tau_A \tau_d}\propto$ L$^{3/2}$, where Alfv\'{e}n time scale is $\tau_A$=4.5 $\mu$s and diffusion time scale is $\tau_d$=316 $\mu$s by assuming the 
current sheet width L=10 cm in laboratory experiment, and then $\tau_{rec, lab}$=38 $\mu$s. The reconnection time scale in solar atmosphere would be, therefore, self-similarly enlarged to 
$\tau_{rec,solar}$=10$^{6-7.5} \tau_{rec,lab} \sim$38-1200 s. This is also comparable to solar observations \citep{nis11, sin11}. 

As mentioned in \S4.1, our experimental results are in the regime of microscopic or marginally-meso scale, so that we cannot directly predict macroscopic MHD phenomena driven by magnetic 
reconnection in solar atmosphere from these experimental results. Nevertheless, the essential reconnection process should be included in the microscopic regime if we suppose Sweet-Parker 
like diffusion region. Thus the microscopic properties, such as reconnection rate, anomalous resistivity, inflow/outflow velocity patterns, observed in this experiment could be compared quantitatively 
with those in the solar observations to see fundamental behaviors of magnetic reconnection.

Numerical simulations would be helpful to connect the microscopic experimental results with the macroscopic solar observations. Global MHD simulation including key reconnection microphysics 
based on the experimental results will be an interesting future work under the framework of laboratory and solar observation collaboration.

\subsection{Wave Mode and Energy Flux}

Magnetic fluctuations of B$_z$ component, i.e. waves, were detected in the current sheet (negative $J_t$ region) during magnetic reconnection. They show multiple frequencies, mainly at 5-6 $\mu$s 
and 10 $\mu$s. Since the measured frequencies are lower than the lower hybrid (LH) frequency ($f_{LH}$=$\sqrt{f_{c,i}f_{c,e}}$=2 MHz, $f_{LH}^{-1}$=0.5 $\mu$s) and rather close to the ion cyclotron 
frequency ($f_{c,i}$=50-150 kHz, $f_{c,i}^{-1}$=6-20 $\mu$s), the measured fluctuations may not be explained by the lower hybrid instability \citep{bal02, car02} nor the modified two-fluid instability 
\citep{ji04}, but by some kind of the drift (kink) instability \citep{zhu96, kuw11} or shear Alfv\'{e}n mode, though it cannot be identified with the current data set. If we consider the oscillation driven by 
the restoring magnetic force due to magnetic reconnection, Alfv\'{e}n time scale determines the oscillation time period, that is, $t_A$=$L$/$V_A$=10 $\mu$s ($L$/25 cm)($V_A$/22 km s$^{-1}$)$^{-1}$, 
assuming the scale length as the half radius of the spheromak (25 cm). This is comparable to the measured frequencies. Furthermore, if we consider Alfv\'{e}n waves generated by magnetic reconnection 
in the solar atmosphere, the wave period would be enlarged to $t_{A,solar}$=10$^{6-7} t_{A,lab}$=10-100 s by considering the scale gap between the laboratory and solar plasmas. This is comparable to 
the solar observations of wave periods 200 s as observed in solar chromospheric jets \citep{nis08, liu09}, an X-ray jet \citep{cir07} and spicules \citep{dep07, oka11}.

The energy fluxes carried by the transverse (Alfv\'{e}n) wave along the guide field in the current sheet in toroidal and axial directions are described by 
\begin{eqnarray}
F_{A, t} = \frac{1}{4\pi}[-\delta B_z \delta v_z B_t -\delta B_z \delta v_t B_z] \\
F_{A, z} = \frac{1}{4\pi}[-\delta B_t \delta v_z B_t -\delta B_t \delta v_t B_z].
\end{eqnarray}
Here we assume that magnetic fluctuations occur only in the 2-dimensional Z-T plane, i.e. $\delta B_r$=$\delta v_r$=0, resulting to the relationships $\delta v_z$=$\frac{\delta B_z}{B_r} v_r$ and $\delta 
v_t$=$\frac{\delta B_t}{B_r} v_r$. Since the fluctuations occur perpendicular to the guide field, we derive $\delta B_z$=-$\frac{B_t}{B_z} \delta B_t$. Furthermore, by using equations such as $\frac{\delta 
B_z}{\delta B_t}$=$\frac{\delta v_z}{\delta v_t}$ and $\frac{v_z}{B_z} =\frac{v_r}{B_r} =\frac{v_t}{B_t}$, we estimate the energy fluxes in the toroidal direction $F_{A,t} \sim$7$\times$10$^6$ erg cm$^{-2}$ 
s$^{-1}$ and in the axial direction $F_{A,z} \sim$3$\times$10$^7$ erg cm$^{-2}$ s$^{-1}$, respectively. This leads to the total energy flux 3.8$\times$10$^7$ erg cm$^{-2}$ s$^{-1}$ and the total energy 
3$\times$10$^7$ erg for the duration of reconnection, 50$\mu$s. This is 1-10\% of the estimated released magnetic energy and comparable to the previous expectations from numerical simulations 
such that 3\% in Yokoyama (1998) and 40\% at maximum in Kigure et al. (2010).

\subsection{Conclusions}

We experimentally investigated fundamental physics of chromospheric jets observed in the solar atmosphere and succeeded in qualitatively reproducing the jets with component magnetic reconnection. 
As an advantage of laboratory experiments, we could directly measure magnetic field strength in 2D plane, plasma (ion) flows, ion temperature, effective magnetic resistivity and high frequency waves in 
association with magnetic reconnection. These measurements are impossible in solar telescope observations and partly in magnetohydrodynamic simulations.

However, on the other hand, qualitatively we found some differences between the laboratory experiment and the solar observations. For example, jet velocity is much smaller than Alfv\'{e}n velocity 
(40\%) in our laboratory experiment, while in solar observations it is comparable to the one. The direction of reconnection jet in laboratory experiment is oblique to the parallel direction to the 
surrounding straight magnetic field lines, while in solar atmosphere it looks parallel to the surrounding vertical magnetic field. The total release energy, the reconnection time scale and the wave 
frequency are also different in laboratory and solar plasmas with several orders.

Here it is useful to consider the differences between microscopic and macroscopic scales, collisionality (collisional and collisionless), which may come from not only the instrumental size but also the 
selection of argon gas, and the degrees of ionization (fully ionized and partially ionized) in solar chromospheric and laboratory plasmas. The reasons for the selection of argon gas are reproductivity and 
emissivity compared with hydrogen, but it makes ion mass and Larmor radius larger than hydrogen which is the majority of gas in the solar atmosphere. We expect that MHD scale physics may play an 
important role in accelerate reconnection jet to the local Alfv\'{e}n speed. This would be impossible to investigate in relatively small scale laboratory experiments compared with the solar atmosphere. 
The scale gap between laboratory and solar plasmas may also explain the differences of the release energy, the reconnection time scale and the wave frequency by the scaling-law. Furthermore, recent 
studies of chromospheric anemone jets suggest that neutral particles in partially ionized plasma, such as chromospheric plasma, may drive fast reconnection \citep{nis11, sin11}, though plasma is fully 
ionized enough during the reconnection event in our current experiment so that it seems that the effect of partial ionization does not appear. The experiment with partially ionized hydrogen plasma 
is an interesting topic in the future work.
\acknowledgements

TS-4 is the two toroidal plasma merging device in Tokyo University. \textit{Hinode} is a Japanese mission developed and launched by ISAS/JAXA, with NAOJ as domestic partner and NASA and STFC 
(UK) as international partners. It is operated by these agencies in co-operation with ESA and NSC (Norway). This work was supported by the JSPS Core-to-Core Program 22001.\\
\\


\begin{figure}[htbp]
\epsscale{.70}
\plotone{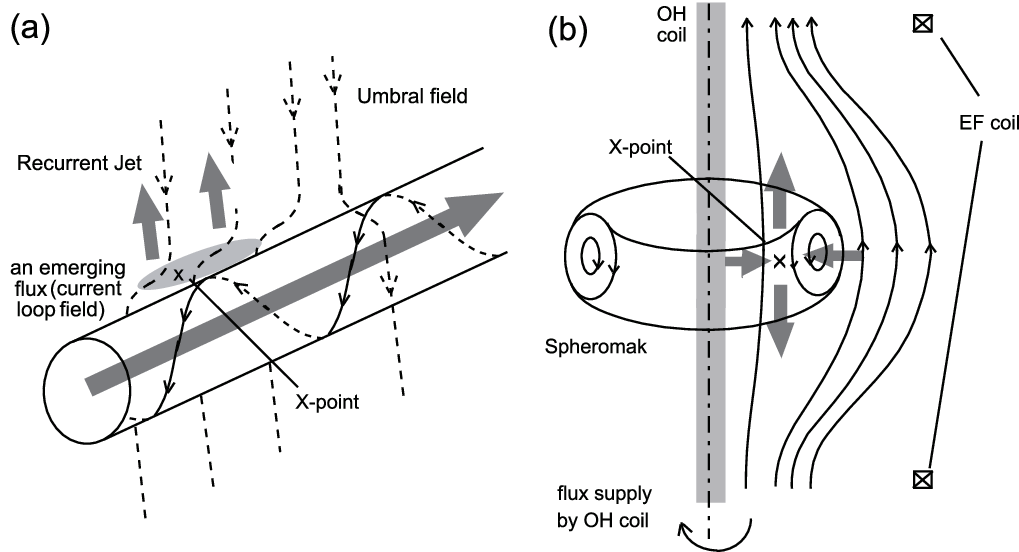}
\caption{Schematic picture of (a) magnetic configuration of the light bridge in the sunspot observed with recurrent plasma ejections compared with the one of (b) the spheromak reconnecting with 
the OH field inside. \label{fig1}}
\end{figure}

\begin{figure}[htbp]
\epsscale{.70}
\plotone{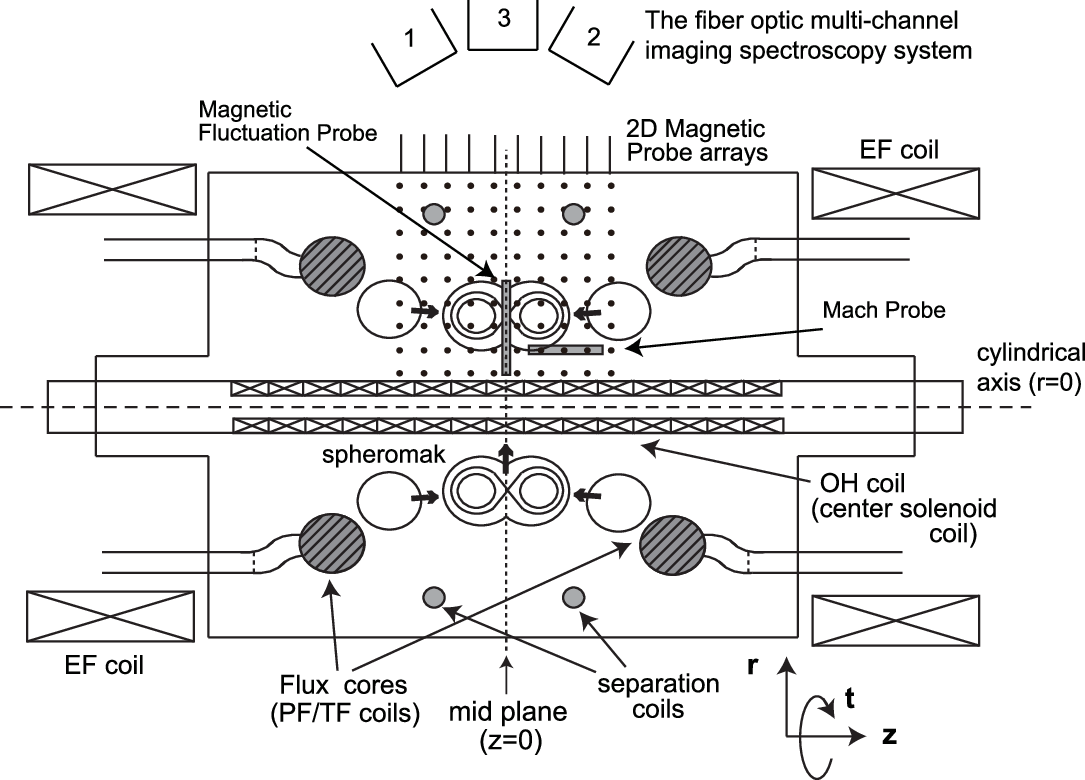}
\caption{TS-4 (Tokyo Spheromak) toroidal plasma merging experiment device. Large and small gray circles show flux cores composed by PF and TF coils and separation coils, respectively. White circles 
show two spheromak plasmas merging on the mid-plane. The direction parallel to the cylindrical axis is Z-direction, based on which other radial and toroidal directions are determined. Dots (or lattice 
points) show the 2D magnetic probe array and the locations of the magnetic fluctuation probe array and the Mach probe are shown with gray colored squares. \label{fig2}}
\end{figure}

\begin{figure}[htbp]
\epsscale{.40}
\plotone{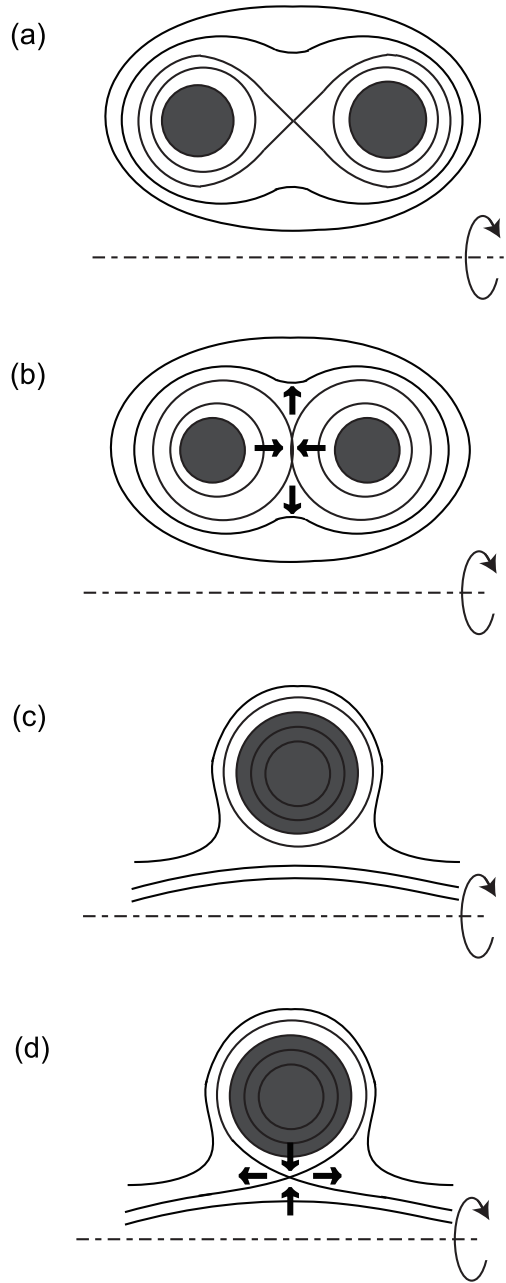}
\caption{Illustration of light bridge reconnection in the double annular plasma configuration in R-Z plane. Thick line shows magnetic field lines and arrows show plasma flows. Gray color means dense 
plasma inside the spheromak. The large spheromak is formed through merging process (c) and reconnects with the new emerging straight magnetic field induced by the center solenoid (OH) coil (d). 
\label{fig3}}
\end{figure}

\begin{figure}[htbp]
\epsscale{.50}
\plotone{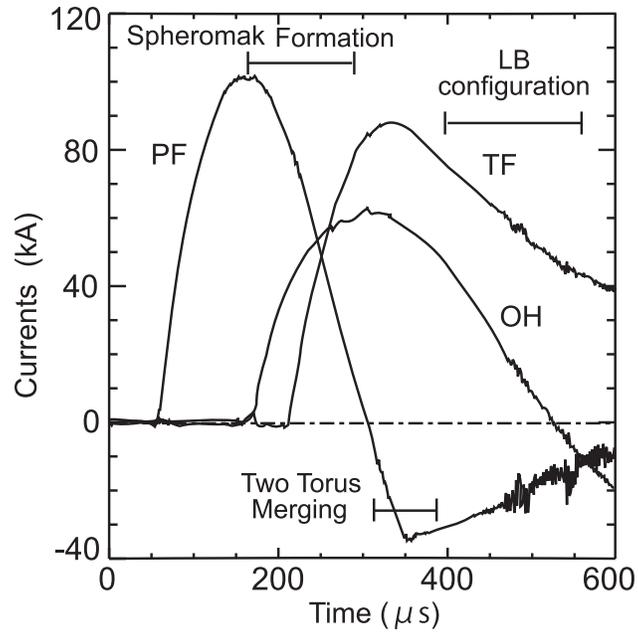}
\caption{Typical current wave forms for TF, PF and OH coils, respectively. \label{fig4}}
\end{figure}

\begin{figure}[htbp]
\epsscale{.70}
\plotone{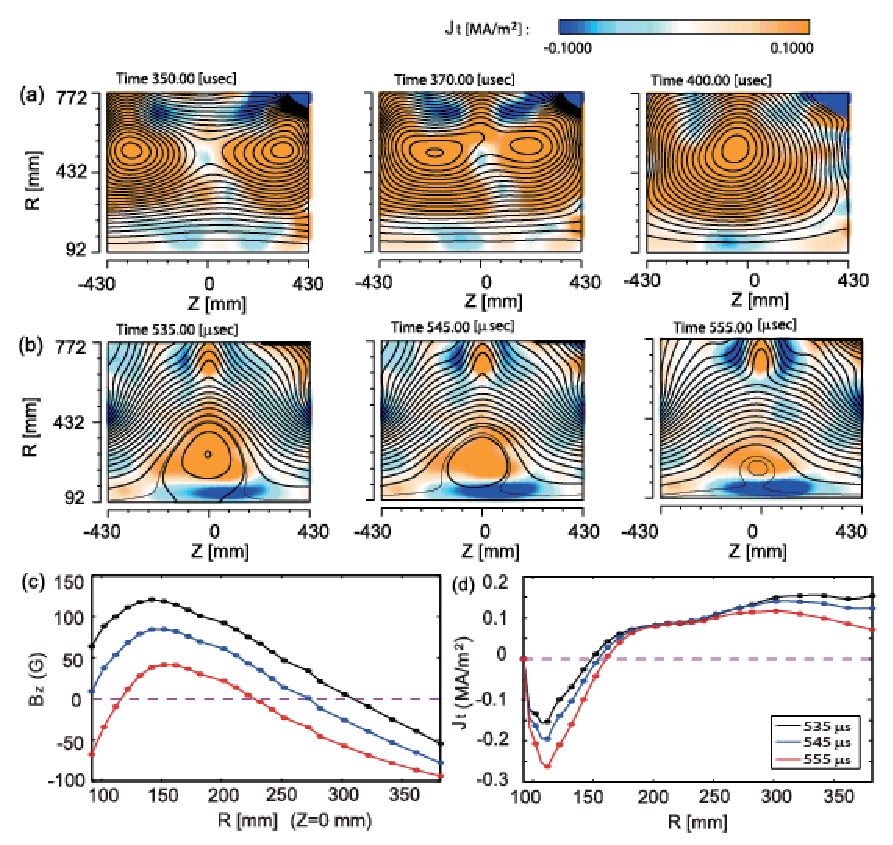}
\caption{Snapshot images of (a) two toroidal argon plasma merging and (b) merging of spheromak and OH field. Thick contours show magnetic field lines at regular intervals and thin contours are 
complementary to emphasize reconnecting field lines. Color bar indicates toroidal current density measured by the 2D magnetic probe array. (c) The radial profile of $B_z$ component and (d) the 
toroidal current $J_t$ on the mid-plane. \label{fig5}}
\end{figure}

\begin{figure}[htbp]
\epsscale{.80}
\plotone{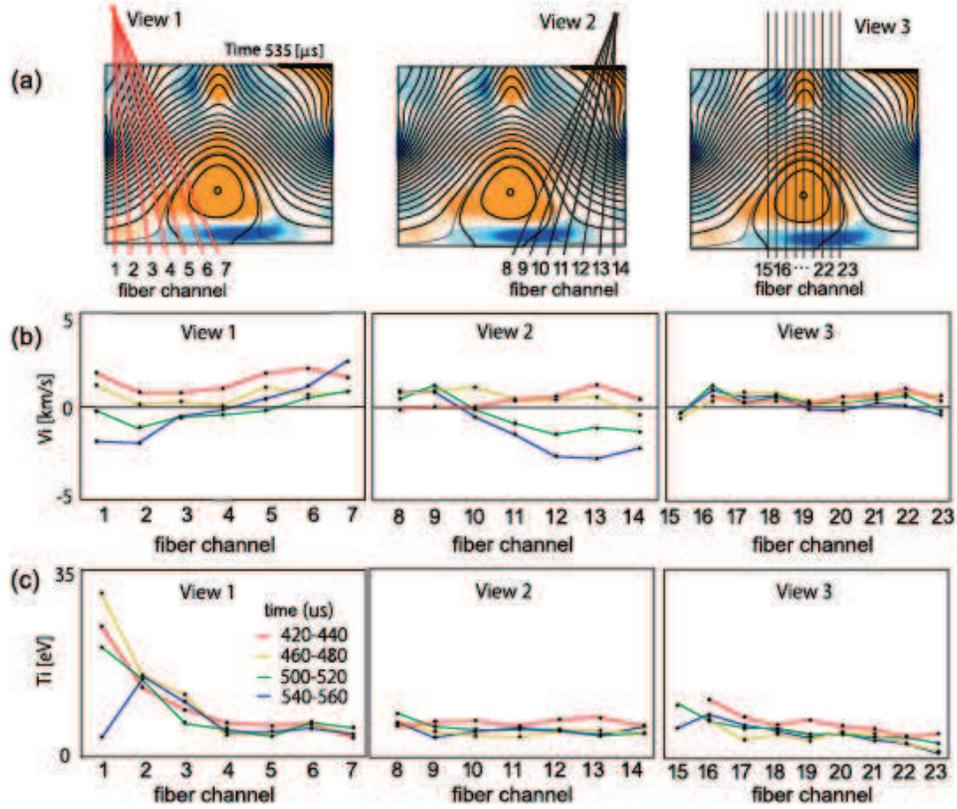}
\caption{(a) The fanned line-of-sight directions of the fiber optic multi-channel imaging spectroscopy system overlaid on a snapshot image of magnetic configuration of the spheromak at 535 $\mu$s 
with toroidal current density in color. (b) 1D axial profiles of ion Doppler velocity in the line-of-sight direction (almost r-direction) with compensation of toroidal velocity component and (c) ion Doppler 
temperature. \label{fig6}}
\end{figure}

\begin{figure}[htbp]
\epsscale{.80}
\plotone{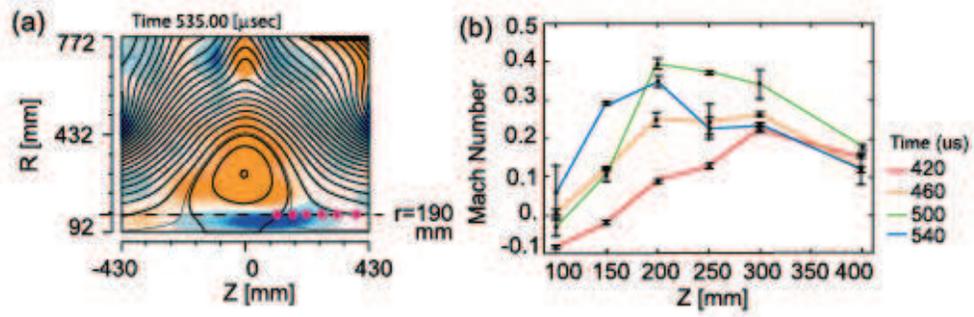}
\caption{(a) The locations of the Mach probe shifted along the current sheet at r=190 mm and z=100, 150, 200, 250, 300 and 400 mm. (b) The axial (Z) profile of ion velocity in axial direction in units 
of Mach number ($C_s \sim$7 km s$^{-1}$) at 420, 460, 500 and 540 $\mu$s. \label{fig7}}
\end{figure}

\begin{figure}[htbp]
\epsscale{.70}
\plotone{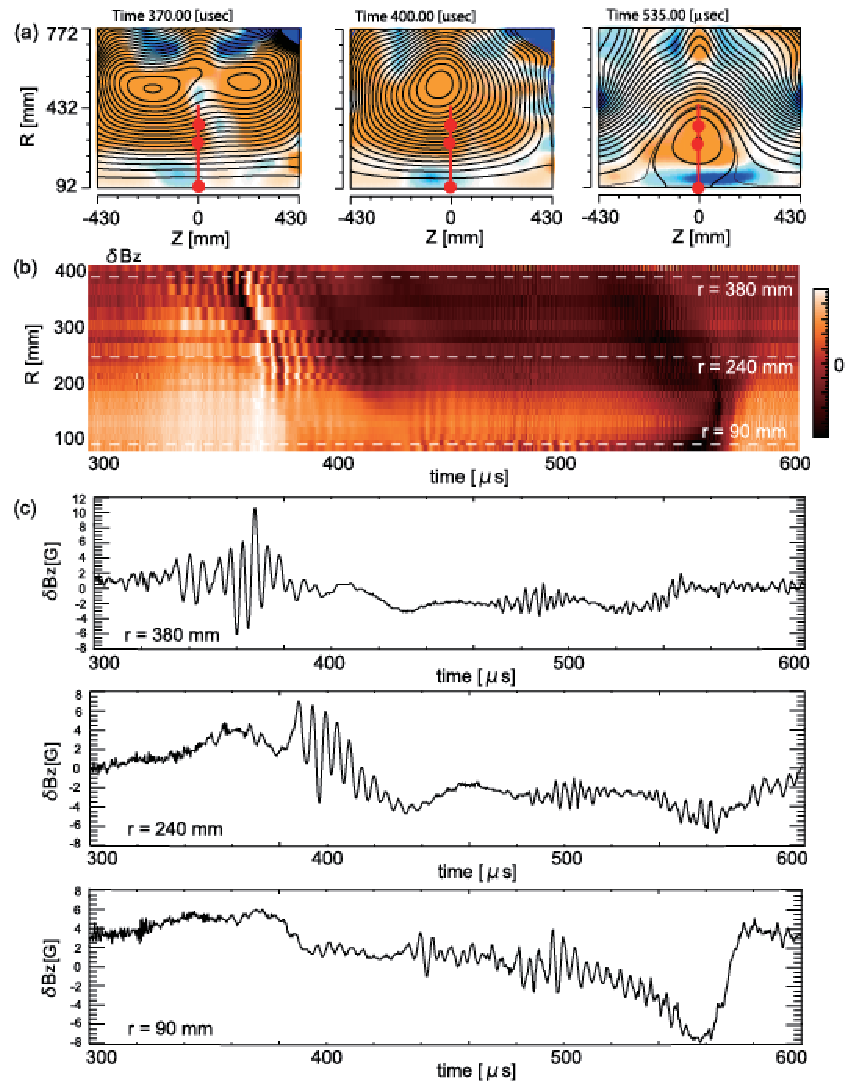}
\caption{(a) The location of the magnetic fluctuation probe array overlaid on snapshot images of plasma merging experiment with poloidal field contours and toroidal current in color. (b) Time slice 
image and (c) time plots of magnetic fluctuations $\delta B_z$ measured by the magnetic fluctuation probe array at $r=$90, 240, 380 mm, whose positions are overlaid on (a) and (b). \label{fig8}}
\end{figure}

\begin{figure}[htbp]
\epsscale{.60}
\plotone{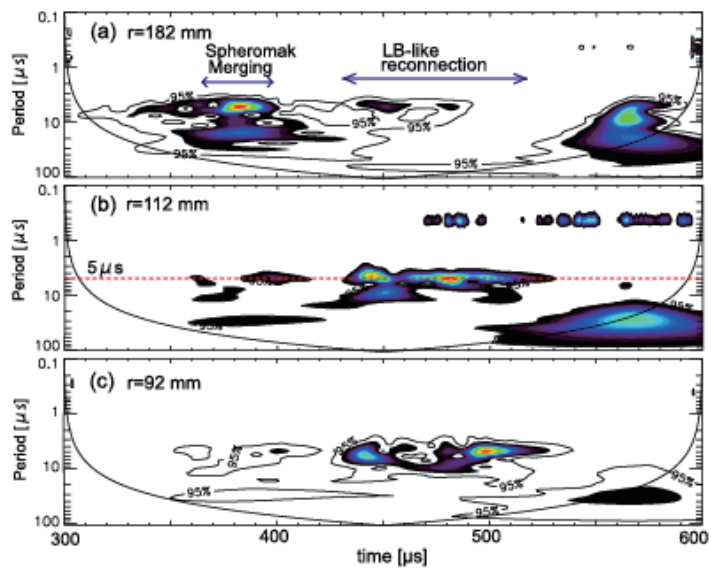}
\caption{Wavelet power spectrum of magnetic fluctuations $\delta B_z$, which is reduced by the smoothed long-term variation larger than 40 $\mu$s from the raw data, measured by the magnetic 
fluctuation probe array on the mid-plane at $r=$92, 112, 182 mm (nearby the OH coil). \label{fig9}}
\end{figure}

\begin{figure}[htbp]
\epsscale{.60}
\plotone{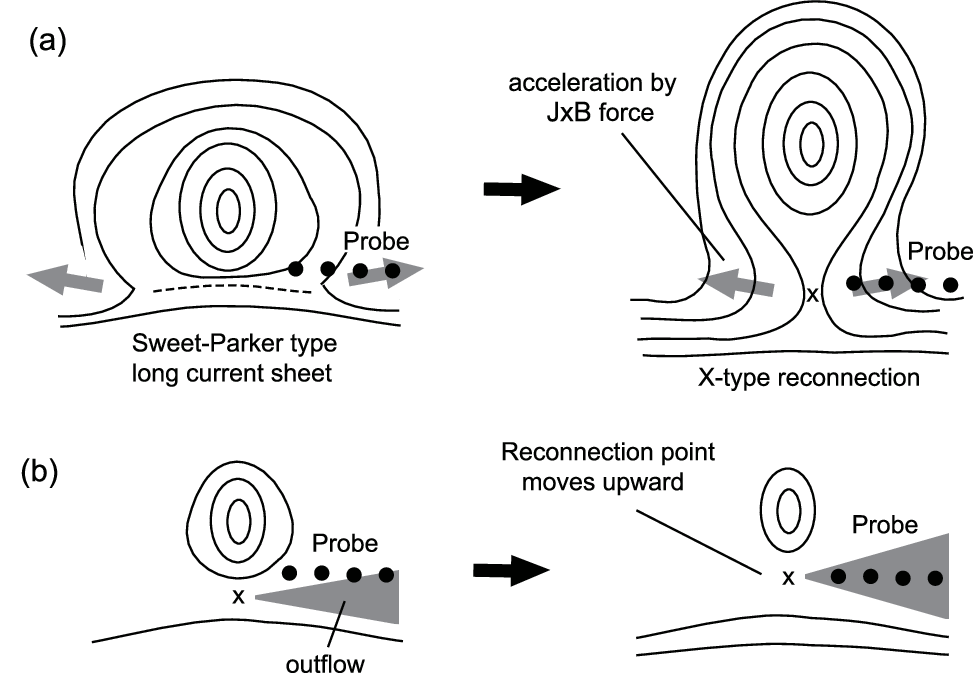}
\caption{Interpretation of accelerated outflow in the inner region of a current sheet: (a) transition from the Sweet-Parker reconnection to X-type reconnection and (b) the movement of the 
reconnection point and outflow up to the detectable area. \label{fig10}}
\end{figure}

\begin{figure}[htbp]
\epsscale{.75}
\plotone{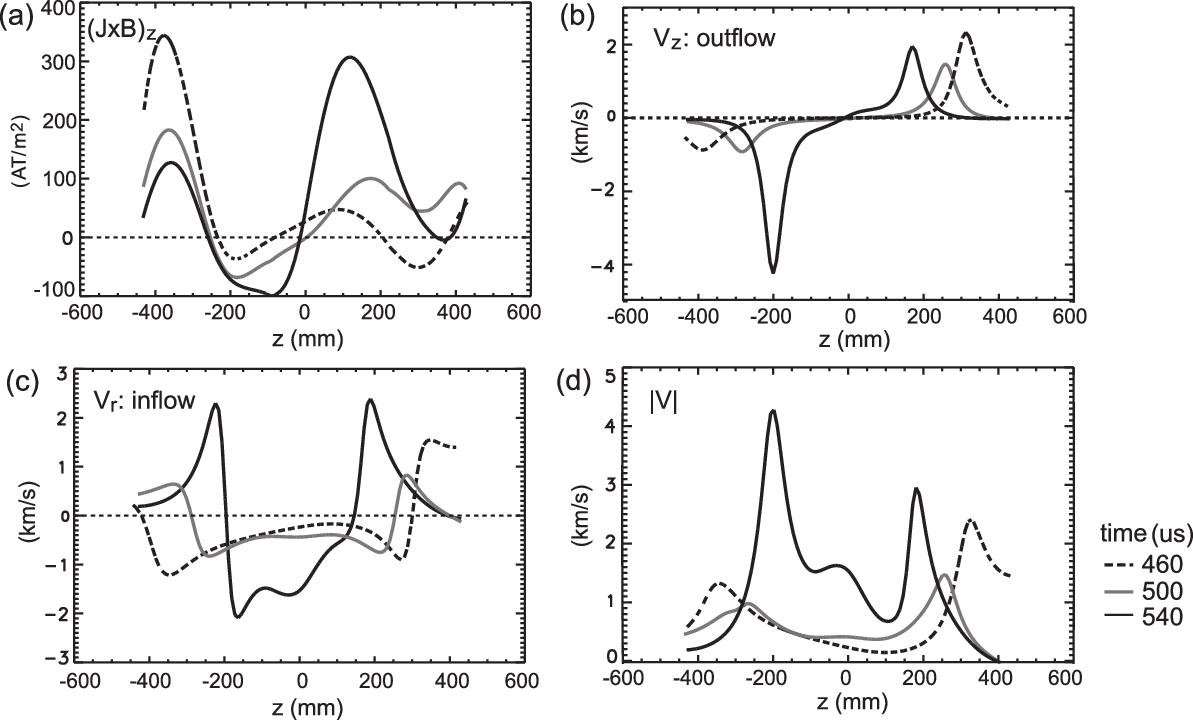}
\caption{The axial (Z) profiles of estimated Lorentz force and plasma velocity perpendicular to the magnetic field lines at r=146 mm: (a) Lorentz force in Z-direction (J$\times$B)$_z$=J$_t$B$_r$, 
(b) axial velocity ${\bf v}_{\perp,z}$=-(E$_t$B$_r$)/$|$B$|^2$ (outflow in Z-direction), (c) radial velocity ${\bf v}_{\perp,r}$=E$_t$B$_z$/$|$B$|^2$ (inflow at the center and outflow in R-direction outside) 
and (d) the absolute value of ion velocity $|{\bf v}_{\perp}|$=E$_t \sqrt{B_t^2+B_r^2}$/$|$B$|^2$. \label{fig11}}
\end{figure}

\begin{figure}[htbp]
\epsscale{.45}
\plotone{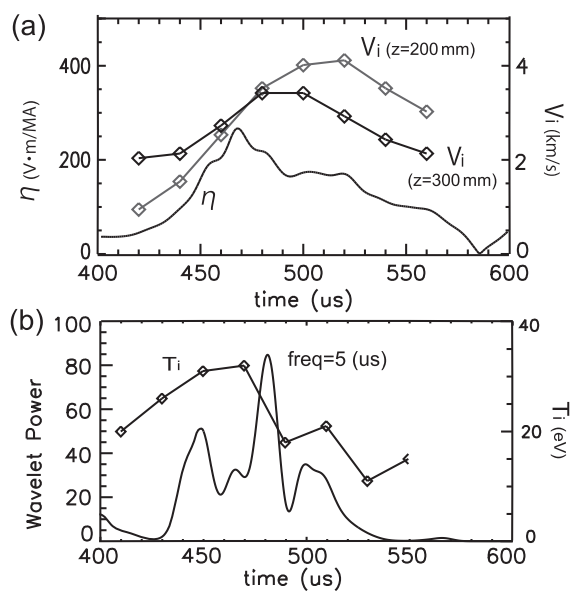}
\caption{(a) Time variation of the effective resistivity $\eta$=E$_t$/J$_t$ and the ion velocities at r=190 mm and z=200 and 300 mm. (b) Time variation of power spectra of magnetic fluctuations with 
5 $\mu$s periods, which is overplotted by ion temperature observed by the Doppler spectroscopy measurement.\label{fig12}}
\end{figure}

\begin{figure}[htbp]
\epsscale{.70}
\plotone{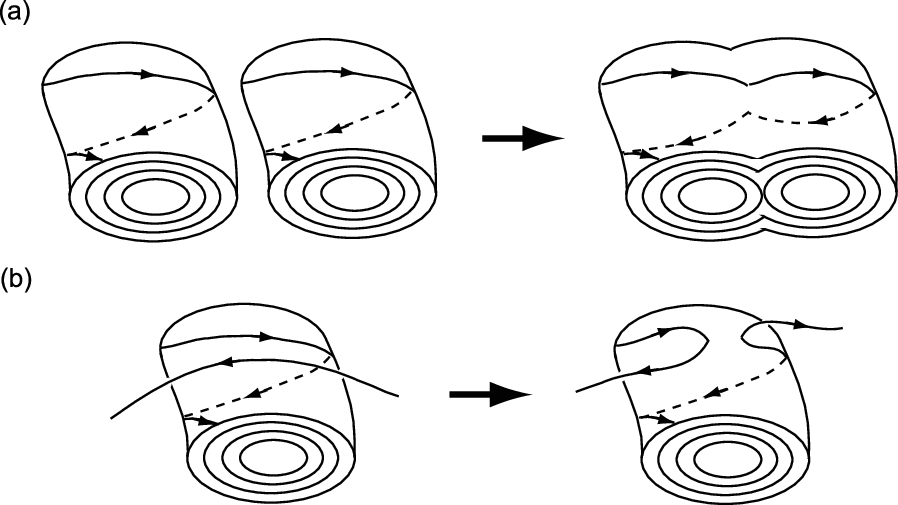}
\caption{Schematic picture of magnetic reconnection (a) between two spheromaks or flux tubes and (b) between a spheromak and the OH field imitating the configuration of an emerging magnetic flux 
rope and the surrounding field. \label{fig13}}
\end{figure}



\clearpage

%
%
\end{document}